\documentclass[11pt]{article}

\usepackage{amsfonts}
\usepackage{amsthm}
\usepackage{amsmath}
\usepackage{enumerate}
\usepackage{graphicx}
\usepackage{color}      
\usepackage{bm}
\usepackage{psfrag}
\usepackage{subfigure}
\usepackage{sidecap}

\def\linespacing#1{\renewcommand{\baselinestretch}{#1}\small\normalsize}

\newtheorem{remark}{Remark}

\newcommand{\vr}{\mathbf{r}}

\newcommand{\vn}{\bm{\nu}}

\newcommand{\vD}{\mathbf{D}}
\newcommand{\vE}{\mathbf{E}}

\newcommand{\vP}{\mathbf{P}}

\newcommand{\vzero}{\mathbf{0}}

\newcommand{\cA}{\mathcal{A}}

\newcommand{\Om}{\Omega}

\newcommand{\Gm}{\Gamma}

\newcommand{\iO}{\int_{\Omega}}

\newcommand{\iG}{\int_{\Gamma}}

\newcommand{\R}{\mathbb{R}}   

\author{Shawn W. Walker}

\begin{document}

\title{On The Correct Thermo-dynamic Potential \\ for Electro-static  Dielectric Energy}

\maketitle

\linespacing{1.0}

\begin{abstract}
Various types of equilibrium processes involve electric fields.  In some cases, the electrical energy appears to be negative (e.g. if the voltage is fixed by an external source).  This paper explains how to derive the correct thermo-dynamic potential for electro-static phenomena, whether the voltage is fixed, or the charge is fixed, or some combination is fixed.  In particular, we explain, in complete detail, why fixing the voltage introduces ``a minus sign'' in the electrical energy.

Two explanations are given.  The first explanation is based on a lumped-parameter argument (i.e. a lumped-capacitor model).  The second explanation uses a distributed parameter model (i.e. a partial differential equation (PDE) model) of a dielectric medium; in this case, we allow for non-linearity and external polarization effects.  Connections with Legendre (duality) transforms are also discussed.
\end{abstract}



\section{Introduction}\label{sec:intro}

The purpose of this paper is to clear up the confusion of why electro-static energy is negative when voltage (potential) is fixed by an external source.  For instance, the potential energy $E$ stored in a linear lumped capacitor of capacitance $C$, with \emph{fixed} total charge $Q$, is
\begin{equation*}
E = \frac{1}{2} C V^2,
\end{equation*}
where $C = Q/V$, and $V$ is the variable voltage potential of the charge.  Since capacitance is always positive, the potential energy is positive.

On the other hand, if the voltage is fixed (by an external source), then the potential energy is usually taken to be
\begin{equation*}
E = -\frac{1}{2} C V^2.
\end{equation*}
In other words, it appears that we simply ``flip'' the sign.  Note that this also occurs when modeling a distributed dielectric medium.

The ``sign flip'' is both confusing (why should potential energy be negative?) and misleading.  The rest of this paper gives a detailed explanation of where this comes from.  Furthermore, we describe how to derive the thermo-dynamically correct, electro-static potential for a distributed dielectric in the presence of non-linearities, external polarization, and non-standard constraints on the charge and potential.  In my experience, this kind of explanation and derivation is very hard to find (if not impossible).  I hope this will help others to better understand how to model electric effects in coupled PDE systems.

\subsection{Background of the Reader}

The reader should have some familiarity with basic physics, e.g. electro-statics, introductory calculus of variations, and basic partial differential equations (PDEs).  Only an undergraduate level of PDE theory is needed, as well as integration by parts (Gauss' divergence theorem) on domains in $\R^3$.  Some knowledge of weak formulations of PDE is beneficial, but not required.  In particular, we do not emphasize issues with function spaces (e.g. the voltage potential $\varphi$ is a function in $H^1(\Omega)$); our main purpose here is in \emph{modeling} issues.  One can easily add the function space aspects if desired.

\subsection{Outline}

Section \ref{sec:lumped_cap} describes a lumped parameter model for the energy stored in a capacitor.  In Section \ref{sec:simple_PDE_arg}, we give a PDE argument for deriving the free energy stored in a dielectric (in a general domain) when the boundary potential is fixed.  Section \ref{sec:general_arg} further justifies the argument in Section \ref{sec:simple_PDE_arg}, as well as generalizing the problem to include extraneous charges and spontaneous (or fixed) polarization.  
We further expand on this modeling approach in Section \ref{sec:fixing_general_bdy_quantity} by considering the situation where one fixes a \emph{non-linear combination} of the boundary potential \emph{and} boundary charge.  We conclude with some remarks in Section \ref{sec:conclusion}.

\section{A (Linear) Lumped Capacitor}\label{sec:lumped_cap}

\subsection{Preliminaries}\label{sec:lumped_cap_prelim}

Suppose two parallel plates (a ``capacitor'') carry an amount of electric charge $Q$.  The presence of the charge \emph{induces} an electric field everywhere, with a net \emph{potential difference} $V$ across the two plate.  The capacitance $C$ of the parallel plates is defined by
\begin{equation}\label{defn:capacitance}
    C = \frac{Q}{V}.
\end{equation}

The amount of electrical energy $E$ (also called dielectric energy) stored in a capacitor is equal to the work done in establishing the electric field between the plates (assuming no losses).  It is given by
\begin{equation}\label{eqn:cap_elec_energy_storage}
    E = \int^Q_0 V(q) \, dq = \int^Q_0 \frac{q}{C} \, dq = \frac{1}{2} \frac{Q^2}{C} = \frac{1}{2} C V^2 = \frac{1}{2} Q V,
\end{equation}
where $C$ is assumed constant while the charge is deposited; note: we used the definition of $C$ \eqref{defn:capacitance}.

\subsection{Application: Perturbing Capacitance}\label{sec:perturb_cap}

In performing experiments, one usually holds certain variables \emph{fixed}. Since electric fields and potentials \emph{originate} from the presence of electric charge, it is reasonable that charge can be used as a control parameter in experiments.  Thus, if we fix $Q$, then $E = \frac{1}{2} \frac{Q^2}{C}$.  Moreover, if the capacitance $C = C(s)$ depends on some parameter $s$ (e.g. by changing the relative permittivity), then the change in potential energy with respect to $s$, \emph{while holding $Q$ fixed}, is given by
\begin{equation}\label{eqn:perturb_cap_value}
    \frac{d E(s)}{d s} = - \frac{1}{2} \frac{Q^2}{C(s)^2} \frac{d C}{d s} = - \frac{1}{2} \frac{d C(s)}{d s} V(s)^2.
\end{equation}

Next, consider the case of fixing the voltage.  It is tempting to posit
\begin{equation}\label{eqn:wrong_thermo_pot}
E = \frac{1}{2} C(s) V^2,
\end{equation}
as the potential energy and differentiate with respect to $s$ to obtain
\begin{equation*}
\frac{d E(s)}{d s} = \frac{1}{2} \frac{d C(s)}{d s} V^2,
\end{equation*}
which has the opposite sign as \eqref{eqn:perturb_cap_value}.  But this is incorrect because experiments show that \cite[Vol. 2]{Feynman_Lectures1964}
\begin{equation*}
	\frac{d E(s)}{d s} = - \frac{1}{2} \frac{d C(s)}{d s} V^2
\end{equation*}
holds true when prescribing the voltage.  The discrepancy is because \eqref{eqn:wrong_thermo_pot} is \textbf{not} the correct thermo-dynamic potential energy when fixing the applied voltage.

\subsection{Fixing the Voltage Difference Across the Plates}

Is it physically possible to exactly \emph{fix} the applied voltage?  \textbf{No!} It is not possible to change the potential without doing something with the charge distributions.  The charges created the potential in the first place.  So fixing the voltage is \emph{not a natural variable} to control in electrical phenomena.

You may object and say that a battery can do this, i.e. a battery is an \emph{idealization} that can take (or give) any amount of charge to keep the potential difference fixed.  This is true, but you must now model the entire system which is composed of the capacitor \emph{and} battery, i.e. the capacitor is no longer an \emph{isolated system}, which was the case when the charge was fixed.


The following sections describe one way to model the coupled battery and capacitor system; this follows the argument in \cite[pg. 119]{Onuki_inbook2005}.


\subsubsection{Lumped Modeling of a Battery}

One can think of a battery as, in fact, an \emph{extremely large} capacitor $C_0$ with an extremely large amount of charge $Q_0$ (i.e. a reservoir of charge).  They are so large, that they dwarf the capacitance and charge of the parallel plate system we considered in Section \ref{sec:lumped_cap_prelim}.  

We now make the following fundamental assumption:
\begin{equation}\label{eqn:battery_assume}
    \text{\textbf{(battery assumption)}} \quad Q_0 \gg Q, \quad C_0 \gg C.
\end{equation}
Next, define the total charge and capacitance of the entire \emph{isolated} system:
\begin{equation}\label{eqn:total_charge_and_cap}
    Q_t = Q_0 + Q, \qquad C_t = C_0 + C,
\end{equation}
where we have assumed (for simplicity) that the ``battery'' and capacitor are connected in parallel (this is not critical though).  Since the system is isolated, the total charge $Q_t$ is \emph{fixed}.  Moreover, we assume $C_0$ is \emph{constant} but $C$ may vary (because of changing the permittivity constant, as in Section \ref{sec:perturb_cap}).

From the assumption, we conclude that
\begin{equation}\label{eqn:approx}
    Q_t \approx Q_0,
\end{equation}
and by the definition of capacitance, we have
\begin{equation*}
    V = \frac{Q_0}{C_0} = \frac{Q}{C} = \frac{Q_t}{C_t}.
\end{equation*}

\subsubsection{Stored Energy Of The Battery}

Next, let us compute (and approximate) the stored electrical energy in the ``battery''.  From \eqref{eqn:cap_elec_energy_storage}, we get
\begin{equation}\label{eqn:battery_energy_deriv_1}
\begin{split}
    E_0 &= \frac{1}{2} \frac{Q^2_0}{C_0} = \frac{1}{2} \frac{(Q_t - Q)^2}{C_0} = \frac{1}{2} \frac{Q^2_t}{C_0} \left( 1 - \frac{Q}{Q_t} \right)^2 \\
    &= \frac{1}{2} \frac{Q^2_t}{C_0} \left( 1 - 2 \frac{Q}{Q_t} + \left( \frac{Q}{Q_t} \right)^2 \right) \\
    &\approx \frac{1}{2} \frac{Q^2_t}{C_0} \left( 1 - 2 \frac{Q}{Q_t} \right) = \frac{1}{2} \frac{Q^2_t}{C_0} - Q \frac{Q_t}{C_0},
\end{split}
\end{equation}
where we dropped $(Q/Q_t)^2$ because it is negligible compared to $2(Q/Q_t)$.  Furthermore, we have
\begin{equation*}
    \frac{Q_t}{C_0} \approx \frac{Q_0}{C_0} = V.
\end{equation*}
So then
\begin{equation}\label{eqn:battery_energy_deriv_2}
\begin{split}
    E_0 &\approx \frac{1}{2} \frac{Q^2_t}{C_0} - Q V,
\end{split}
\end{equation}
where the first term is a \emph{fixed constant}.

\subsubsection{Total Energy}

The energy of the small capacitor, is of course $E = \frac{Q V}{2}$ by \eqref{eqn:cap_elec_energy_storage}.  Thus, the total energy of the isolated system is the sum:
\begin{equation}\label{eqn:total_energy}
\begin{split}
    E_t &= E_0 + E \approx \frac{1}{2} \frac{Q^2_t}{C_0} - Q V + \frac{1}{2} Q V = \frac{1}{2} \frac{Q^2_t}{C_0} - \frac{1}{2} Q V,
\end{split}
\end{equation}
where the first term is a fixed constant and $V$ is \emph{approximately} constant.  For the purposes of energy perturbation (or finding energy minimizers), one can obviously drop the first term.  This gives the following \emph{thermodynamic potential} when $V$ is fixed:
\begin{equation}\label{eqn:thermo_potential}
    \widetilde{E}_t := -\frac{1}{2} C V^2,
\end{equation}
where we used the definition of the capacitance \eqref{defn:capacitance}.  Note the minus sign!  Thus, assuming $C = C(s)$ and following the argument in Section \ref{sec:perturb_cap}, differentiating with respect to $s$ gives
\begin{equation*}
 \frac{d}{ds} \widetilde{E}_t (s) = -\frac{1}{2} \frac{d C(s)}{ds} V^2,
\end{equation*}
which has the correct sign (as based on experiments \cite[Vol. 2]{Feynman_Lectures1964}).  We remark that an alternative derivation of the energy (with the minus sign) is given in \cite[Vol. 2]{Feynman_Lectures1964}.

\subsection{Legendre (Duality) Transform}

The above derivation is rather involved.  Typically, one uses a  Legendre transform to hide these details.  One can think of $Q$ and $V$ as dual variables; their product has units of energy.

When $Q$ is the control variable, the energy stored in the single capacitor is simply $E = \frac{Q^2}{2 C}$.  In order to shift the control variable to $V$, we define the \emph{conjugate energy} by a (non-standard) Legendre transform, i.e.
\begin{equation}\label{eqn:Legendre_trans_electrical_energy}
    E^* = E - Q V = \frac{Q^2}{2 C} - QV = \frac{Q V}{2} - Q V = -\frac{1}{2} Q V = -\frac{1}{2} C V^2,
\end{equation}
which is the correct thermodynamic potential to use when $V$ is fixed.  This lets one ignore the battery argument.  Note that $E^*$ is convex in $Q$ since $V$ is fixed.  Indeed, for a given $V$, the isolated system finds a minimum $Q$ at equilibrium.  This is discussed more in Sections \ref{sec:simple_PDE_arg} and \ref{sec:general_arg}.

\begin{remark}
Legendre transforms are common in thermodynamics.  The purpose of making the duality transform is to avoid explicitly dealing with the ``reservoir'' argument.  Some variables are convenient to think about as control variables (e.g. the voltage) because they are easy to measure.  But in reality, they are not so easy to control or fix in an isolated system.  One has to model a reservoir to make it work.
\end{remark}

The same arguments hold if the small capacitor is replaced by a distributed system (e.g. Laplace's equation, an integral of the Dirichlet energy $\int_\Omega \varepsilon |\nabla V|^2$, etc.).  The battery part of the total energy can still be treated as a \emph{lumped} object.  One needs to use the fact that $Q$ is related to the surface charge density on the small capacitor and identify it with the Neumann data $\vn \cdot (\varepsilon \nabla V)$ (see \cite{Landau_Book1960} for an explanation of this).  We explain this in more detail in Section \ref{sec:simple_PDE_arg}.

\section{A Simple PDE Argument}\label{sec:simple_PDE_arg}

The previous explanation is adequate, but is not very general. Here we consider a distributed system and model the electric field with a PDE.

\subsection{Energy of the Dielectric}

Let $\Om$ be the domain of the dielectric with permittivity $\varepsilon$.  The potential is denoted by $\varphi : \Om \to \R$.  The energy of the dielectric, with no other extraneous charges, is given by
\begin{equation}\label{eqn:dielectric_energy}
  J(\varphi) = \frac{1}{2} \iO \varepsilon |\nabla \varphi|^2,
\end{equation}
where $\varphi$ \emph{solves the PDE} \eqref{eqn:dielectric_energy_Neumann_E-L} (fixed boundary charge).  Note: we derive this in Section \ref{sec:fixed_bdy_charge_case}.  For now, we take \eqref{eqn:dielectric_energy} as given from \emph{physics}.

The next section further discusses the fixed boundary charge case.  In Section \ref{sec:dirichlet_potential}, we describe how the energy changes when fixing the potential on the boundary of $\Om$.  

\subsection{Neumann}\label{sec:Neumann}

Let us consider the case where some charge distribution $q$ is fixed on the boundary: $\Gm = \partial \Om$.
%
The equilibrium potential (up to an arbitrary constant) satisfies
\begin{equation}\label{eqn:dielectric_energy_Neumann_E-L}
\begin{split}
  -\nabla \cdot (\varepsilon \nabla \varphi) &= 0, \text{ in } \Om, \\
  \vn \cdot (\varepsilon \nabla \varphi) &= q, \text{ on } \Gm,
\end{split}
\end{equation}
provided $\iG q = 0$ (otherwise, there is no equilibrium solution).  In what follows, for simplicity, we assume $\Om$ is surrounded by a perfect conductor.

Let $\varphi^* = \varphi^*(\varepsilon)$ be the equilibrium solution, which depends on $\varepsilon$.  We \emph{assume} that changes in $\varepsilon$ will not affect $q$, i.e. $q$ is fixed irrespective of the material in $\Om$.  It is presumed that the charges are permanently attached to material objects which are then fixed.  The case of $q=0$, i.e. insulating boundary conditions, is a standard setting (of course, in this case, the equilibrium solution $\varphi^*$ is an arbitrary constant.)

Hence, this is an \textbf{isolated system} with respect to changes in $\varepsilon$, because the outside world is shielded by the perfect conductor.  Note that the constant (or mean value part) of $\varphi^*$ \emph{does depend on the environment outside} $\Om$, i.e. the conductor will have some constant potential.  But this has \emph{no effect} on the energy.

The free energy in this case is given by \eqref{eqn:dielectric_energy}.  Thus, if we want to know how the free energy changes with respect to $\varepsilon$, then we simply compute
\begin{equation}\label{eqn:dielectric_energy_Neumann_perturb}
  \delta_{\varepsilon} J(\varphi^*) = \delta_{\varepsilon} \left( \frac{1}{2} \iO \varepsilon |\nabla \varphi^*|^2 \right),
\end{equation}
where we note that $\varphi^* = \varphi^*(\varepsilon)$.  So, \eqref{eqn:dielectric_energy_Neumann_perturb} is evaluated by plugging in $\delta_{\varepsilon} \varphi^*$, which is the solution of a PDE obtained by differentiating the PDE \eqref{eqn:dielectric_energy_Neumann_E-L} with respect to $\varepsilon$.

\subsection{Dirichlet}\label{sec:dirichlet_potential}

Let us consider the case where the potential $\varphi$ is fixed on $\Gm$, i.e. $\varphi = g$ on $\Gm$ and $g$ is fixed.
The equilibrium potential satisfies
\begin{equation}\label{eqn:dielectric_energy_Dirichlet_E-L}
\begin{split}
  -\nabla \cdot (\varepsilon \nabla \varphi) &= 0, \text{ in } \Om, \\
  \varphi &= g, \text{ on } \Gm.
\end{split}
\end{equation}
This problem is essentially equivalent to \eqref{eqn:dielectric_energy_Neumann_E-L} if one invokes the Dirichlet-to-Neumann map. The Neumann data $\vn \cdot \varepsilon \nabla \varphi$ can be interpreted as a surface charge density.  According to electric theory, this is really what creates the electric field inside $\Om$.

\subsubsection{Work Done By The Battery}

Let $\varphi^* = \varphi^*(\varepsilon)$ be the equilibrium solution, which depends on $\varepsilon$.  We \emph{assume} that changes in $\varepsilon$ will not affect $g$, i.e. $\varphi = g$ is fixed irrespective of the material in $\Om$.  What does this mean or imply?

Changing $\varepsilon$ \emph{anywhere} in $\Om$ will change $\vn \cdot \varepsilon \nabla \varphi$ on $\Gm$, even if $\varepsilon$ is fixed on $\Gm$.  This is equivalent to the local charge density $q \equiv \vn \cdot \varepsilon \nabla \varphi$ changing.  The charge density may become less dense in some areas but more dense elsewhere.  Since we assume the potential on the boundary is fixed, this cannot happen without some assistance from the outside world.  In particular, the voltage source (e.g. the battery) must do work to move (and supply) these charges around while keeping the potential fixed.  This amount of work is given by
\begin{equation}\label{eqn:battery_work}
  \text{work done by the battery:  } = - \iG g \delta_{\varepsilon} q = - \iG g \delta_{\varepsilon} (\vn \cdot \varepsilon \nabla \varphi),
\end{equation}
i.e. multiply the change in charge by its potential.  (If you are wondering about the minus sign, then read Section \ref{sec:general_arg}.)

Therefore, the dielectric domain is \textbf{not} an isolated system with respect to changes in $\varepsilon$.  The correct isolated system consists of the dielectric domain \emph{and} the voltage source.

\subsubsection{Change in Free Energy}

So, if we want to know how the free energy changes with respect to $\varepsilon$, then we must account for the work done by the battery:
\begin{equation}\label{eqn:dielectric_energy_Dirichlet_perturb}
\begin{split}
  \delta_{\varepsilon} \text{ (free energy) } &= \delta_{\varepsilon} J(\varphi^*) - \iG g \delta_{\varepsilon} (\vn \cdot \varepsilon \nabla \varphi^*) \\
  &= \delta_{\varepsilon} \left( \frac{1}{2} \iO \varepsilon |\nabla \varphi^*|^2 \right) - \iG g \delta_{\varepsilon} (\vn \cdot \varepsilon \nabla \varphi^*) \\
  &= \delta_{\varepsilon} \left\{ \left( \frac{1}{2} \iO \varepsilon |\nabla \varphi^*|^2 \right) - \iG g (\vn \cdot \varepsilon \nabla \varphi^*) \right\},
\end{split}
\end{equation}
where $g$ is fixed. Note the difference with \eqref{eqn:dielectric_energy_Neumann_perturb}.

Therefore, the ``effective'' energy in this context is
\begin{equation}\label{eqn:dielectric_energy_Dirichlet_effective}
\begin{split}
  \widetilde{J}(\varphi^*) &= \left( \frac{1}{2} \iO \varepsilon |\nabla \varphi^*|^2 \right) - \iG g (\vn \cdot \varepsilon \nabla \varphi^*) = -\frac{1}{2} \iO \varepsilon |\nabla \varphi^*|^2,
\end{split}
\end{equation}
which follows by using the PDE \eqref{eqn:dielectric_energy_Dirichlet_E-L} and applying integration by parts.  Eureka!  There is the minus sign.  The choice of the sign in \eqref{eqn:battery_work} can be reasoned on physical grounds (or read the argument in Section \ref{sec:general_arg}).  Moreover, one can interpret \eqref{eqn:dielectric_energy_Dirichlet_effective} as a (non-standard) Legendre transform.  The next section gives a more general, first principles, argument for deriving the free energy.

\section{A More General Argument}\label{sec:general_arg}

This explanation starts from basics in electromagnetic field theory.

\subsection{Fundamental Equations}

We assume we are in a slowly time-varying regime, so the first basic equations are
\begin{equation}\label{eqn:maxwell_static_eqns_1}
\begin{split}
  \nabla \times \vE &= \vzero, \\
  \nabla \cdot \vD &= \rho,
\end{split}
\end{equation}
where $\rho$ is the \emph{free} charge density; $\vE$ is the electric field, and $\vD$ is the ``displacement'' field.  We must connect these two quantities by a constitutive relation:
\begin{equation}\label{eqn:vD_vE_relation}
  \vD = \varepsilon_0 \varepsilon_r \vE + \vP,
\end{equation}
where $\vP$ could be an additional polarization vector coming from some other source (take it as given).

\begin{remark}
Suppose we have a dielectric domain surrounded by a perfect conductor.  Consider a surface element $S$ on the boundary of the dielectric domain, where the normal vector $\vn$ points into the conductor. Let $S^{\pm}$ bound a thin volume region $V$ around $S$, i.e. $S^{\pm}$ are infinitesimally close to $S$.  Since $\vE$ and $\vP$ vanish inside the conductor, we have that
\begin{equation*}
- \int_{S} \vD \cdot \vn = \int_{S^{\pm}} \vD \cdot \vn_{S^{\pm}} = \int_{V} \nabla \cdot \vD = \int_{V} \rho = \int_{S} \sigma,
\end{equation*}
where we assume $\rho = \delta_{S} \sigma$, $\sigma$ is the surface charge density, and $\delta_{S}$ is a Dirac mass on $S$.  (Note, we are taking a limit in the thinness.)  Therefore, since $S$ is arbitrary, $-\vD \cdot \vn = \sigma$.
\end{remark}

Since $\nabla \times \vE = \vzero$, there exists $\varphi$ such that $\vE = - \nabla \varphi$.  Then, \eqref{eqn:maxwell_static_eqns_1}, \eqref{eqn:vD_vE_relation} implies
\begin{equation}\label{eqn:electro-static_eqn_temp_1}
\begin{split}
  -\nabla \cdot (\varepsilon_0 \varepsilon_r \nabla \varphi - \vP) &= \rho, \text{ in } \Om,
\end{split}
\end{equation}
on the dielectric domain $\Om$; again note that $\rho$ is the free charge density.  Moreover, we have $\iO \rho = 0$, because we assume that there are only internal dipole moments.

Now suppose we are in a situation where we can control the distribution of charge on the boundary and keep the system disconnected from anything else.  In this case,
\begin{equation}\label{eqn:electro-static_eqn_fixed_charge}
\begin{split}
  \nabla \cdot \vD \equiv -\nabla \cdot (\varepsilon_0 \varepsilon_r \nabla \varphi - \vP) &= \rho_B, \text{ in } \Om, \\
  -\vn \cdot \vD \equiv \vn \cdot \varepsilon_0 \varepsilon_r \nabla \varphi - \vn \cdot \vP &= \sigma, \text{ on } \Gm,
\end{split}
\end{equation}
where $\rho = \rho_B + \delta_{\Gm} \sigma$, $\sigma$ is a known surface free charge density, and $\rho_B$ is the known bulk charge density which does not include a singular (concentrated) term on the boundary.  This gives
\begin{equation}\label{eqn:compatible_charge}
  \iG \sigma = -\iG \vn \cdot \vD = - \iO \nabla \cdot \vD = - \iO \rho_B \quad \Rightarrow \quad \iO \rho_B + \iG \sigma = 0,
\end{equation}
i.e. we have compatibility.

We emphasize that $\vP$ must be considered in the distribution of charges (since it is from an external source) and is also fixed.  In other words,
\begin{equation}\label{eqn:electro-static_eqn_fixed_charge_final}
\begin{split}
  -\nabla \cdot (\varepsilon_0 \varepsilon_r \nabla \varphi) &= \rho_B - \nabla \cdot \vP, \text{ in } \Om, \\
   \vn \cdot \varepsilon_0 \varepsilon_r \nabla \varphi &= \sigma + \vn \cdot \vP, \text{ on } \Gm,
\end{split}
\end{equation}
Similar to \eqref{eqn:compatible_charge}, we must have $\iO \rho_B + \iG \sigma = 0$.  Since $\vP$ is fixed, the potential is a function of $\rho_B$ and $\sigma$: $\varphi = \varphi (\rho_B,\sigma)$.

\textbf{Example.} Suppose there is \textbf{no} free charge (except for the external spontaneous polarization).  Then,
\begin{equation}\label{eqn:electro-static_eqn_fixed_ZERO_charge}
\begin{split}
  -\nabla \cdot (\varepsilon_0 \varepsilon_r \nabla \varphi) &= -\nabla \cdot \vP, \text{ in } \Om, \\
  \vn \cdot \varepsilon_0 \varepsilon_r \nabla \varphi &= \vn \cdot \vP, \text{ on } \Gm,
\end{split}
\end{equation}
i.e. $\varphi$ is completely determined (up to a constant) by $\vP$.  If $\vP = \vzero$, then $\varphi$ is an arbitrary constant.  Clearly, the energy should be zero in this case.

\subsection{Free Energy in the Fixed Boundary Charge Case}\label{sec:fixed_bdy_charge_case}

Recall that the energy of $N$ discrete charges $\{ q_i \}_{i=1}^N$ is given by
\begin{equation*}
  U = \frac{1}{2} \sum_{i=1}^N q_i \Phi(\vr_i),
\end{equation*}
where $\Phi$ is the global potential induced by all the charges and $\vr_i$ is the location of charge $q_i$.  Note the $1/2$ is because of ``double counting.''

We can generalize this to the continuum setting by
\begin{equation}\label{eqn:electro-static_eqn_energy_fixed_charge_init}
\begin{split}
  U = \frac{1}{2} \iO \rho \varphi &= \frac{1}{2} \left[ \iO \tilde{\rho}_B \varphi + \iG \tilde{\sigma} \varphi \right] \\
  &= \frac{1}{2} \left[ \iO (\rho_B - \nabla \cdot \vP) \varphi + \iG (\sigma + \vn \cdot \vP) \varphi \right],
\end{split}
\end{equation}
where $\tilde{\rho}_B = \rho_B - \nabla \cdot \vP$ is the effective bulk charge and $\tilde{\sigma} = \sigma + \vn \cdot \vP$ is the effective surface charge.

We further \emph{note that $\varphi$ solves} \eqref{eqn:electro-static_eqn_fixed_charge_final}; thus, we can further manipulate \eqref{eqn:electro-static_eqn_energy_fixed_charge_init} with \eqref{eqn:electro-static_eqn_fixed_charge_final} using integration by parts:
\begin{equation}\label{eqn:electro-static_eqn_energy_fixed_charge_final}
\begin{split}
  U &= \frac{1}{2} \left[ \iO -\nabla \cdot (\varepsilon_0 \varepsilon_r \nabla \varphi) \varphi + \iG (\vn \cdot \varepsilon_0 \varepsilon_r \nabla \varphi) \varphi \right] = \frac{1}{2} \iO (\varepsilon_0 \varepsilon_r \nabla \varphi) \cdot \nabla \varphi \\
  &= \frac{1}{2} \iO \varepsilon_0 \varepsilon_r |\nabla \varphi|^2
\end{split}
\end{equation}
which is the ``standard'' internal dielectric energy in $\Om$.  This is the thermo-dynamic potential (i.e. free energy) to use when controlling the boundary charge on a dielectric with a conductor surrounding it.

\begin{remark}\label{rem:Neumann_dependence_thermo_potential}
Since $\varphi$ solves \eqref{eqn:electro-static_eqn_fixed_charge_final}, then $\varphi = \varphi (\rho_B,\sigma)$.  Hence, the free energy depends on $\rho_B$ and $\sigma$: $U = U(\rho_B,\sigma)$ when controlling the charge.  If $\rho_B$ never changes, then $U = U(\sigma)$.  Note that $U$ also (implicitly) depends on $\Om$ and $\varepsilon_r$.
\end{remark}

\subsection{Free Energy When Fixing the Boundary Potential}\label{sec:free_energy_fixed_bdy_potential}

In Section \ref{sec:fixed_bdy_charge_case}, the internal energy has $\sigma$ as an \emph{independent} control variable (note: that $\vP$ is taken as fixed).  For simplicity, we take $\rho_B$ as a given fixed function.

We now need to find the potential energy function that has $\varphi |_{\Gm}$ (i.e. the boundary voltage) as an independent control variable.  In other words, how does controlling the boundary voltage affect the free energy?

\subsubsection{Fixing the Boundary Potential}\label{sec:fixing_bdy_potential}

As noted earlier, fixing the boundary voltage is not possible without changing the isolated system.  Something must be added to the original system.  So, we add a voltage source (a reservoir of charge at a fixed potential $g$), which effectively adds another term to the energy \eqref{eqn:electro-static_eqn_energy_fixed_charge_final}:
\begin{equation}\label{eqn:modified_energy_for_fixed_pot}
   \cA(\sigma,g) = U(\sigma) - \iG \sigma g
\end{equation}
where $\iG \sigma g$ accounts for the change in energy due to the change in charge on the boundary. 
Note: we keep $\rho_{B}$ and $\vP$ fixed. Again, we invoke the Legendre transform.

How do we know $- \iG \sigma g$ is the ``correct thing to add?''  We comment on this in Remark \ref{rem:choice_legendre_transform}.

\subsubsection{Finding a New Equilibrium}\label{sec:re_equilibrate}

Now that we have connected the voltage source, we let the new isolated system find its equilibrium given that $g$ is fixed.  In other words, the surface charge density $\sigma$ will obtain a value dependent on $g$ that achieves a stable equilibrium for $\cA$.  Thus, the new thermo-dynamic potential is
\begin{equation}\label{eqn:electro-static_eqn_energy_fixed_pot}
\begin{split}
\widehat{U}(g) &= \min_{\sigma} \cA(\sigma,g),
\end{split}
\end{equation}
where we recall \eqref{eqn:electro-static_eqn_energy_fixed_charge_final} and Remark \ref{rem:Neumann_dependence_thermo_potential}.  In particular, let $\varphi = \varphi(\sigma)$ solve \eqref{eqn:electro-static_eqn_fixed_charge_final} (note that we keep $\rho_B$ and $\vP$ fixed).

Let $\varphi' \equiv \varphi'(\sigma) := \delta_{\sigma} \varphi (\sigma) \cdot \xi$ ($\xi$ is the perturbation of $\sigma$); hence, $\varphi'$ solves
\begin{equation}\label{eqn:electro-static_eqn_fixed_charge_perturb}
\begin{split}
-\nabla \cdot (\varepsilon_0 \varepsilon_r \nabla \varphi') &= 0, \text{ in } \Om, \\
\vn \cdot \varepsilon_0 \varepsilon_r \nabla \varphi' &= \xi, \text{ on } \Gm.
\end{split}
\end{equation}
Therefore, at equilibrium, for a given $g$, we must have
\begin{equation}\label{eqn:electro-static_eqn_energy_fixed_pot_min_prop}
\begin{split}
\delta_{\sigma} \cA(\sigma; \xi) = 0, \quad \forall \xi.
\end{split}
\end{equation}
Note that this equilibrium point is indeed a minimizer because $U(\sigma)$ \emph{is convex} in $\sigma$ and the additional term is only linear in $\sigma$.  Let $\hat{\sigma}$ be the unique solution of \eqref{eqn:electro-static_eqn_energy_fixed_pot_min_prop}.

Computing \eqref{eqn:electro-static_eqn_energy_fixed_pot_min_prop} explicitly, we get
\begin{equation}\label{eqn:electro-static_eqn_energy_fixed_pot_E-L_pt_1}
\begin{split}
   \iO \varepsilon_0 \varepsilon_r \nabla \varphi (\hat{\sigma}) \cdot \nabla \varphi' - \iG \xi g  = 0, \quad \forall \xi.
\end{split}
\end{equation}
Using \eqref{eqn:electro-static_eqn_fixed_charge_perturb} and integrating by parts, we get
\begin{equation}\label{eqn:electro-static_eqn_energy_fixed_pot_E-L_pt_2}
\begin{split}
   \iO \underbrace{-\nabla \cdot \left( \varepsilon_0 \varepsilon_r \nabla \varphi' \right)}_{=0} \varphi (\hat{\sigma}) + \iG \underbrace{\vn \cdot \left( \varepsilon_0 \varepsilon_r \nabla \varphi' \right)}_{=\xi} \varphi (\hat{\sigma}) - \iG \xi g = 0, \quad \forall \xi,
\end{split}
\end{equation}
and so
\begin{equation}\label{eqn:electro-static_eqn_energy_fixed_pot_E-L_pt_3}
\begin{split}
  \iG \xi (\varphi (\hat{\sigma}) - g) = 0, \quad \forall \xi,
\end{split}
\end{equation}
which implies that $\varphi (\hat{\sigma}) = g$ on $\Gm$, i.e. the solution $\varphi (\hat{\sigma})$ of \eqref{eqn:electro-static_eqn_fixed_charge_final} that corresponds to the minimizer $\hat{\sigma}$ in \eqref{eqn:electro-static_eqn_energy_fixed_pot} must also solve
\begin{equation}\label{eqn:electro-static_eqn_fixed_pot_final}
\begin{split}
-\nabla \cdot (\varepsilon_0 \varepsilon_r \nabla \varphi) &= \rho_B - \nabla \cdot \vP, \text{ in } \Om, \\
\varphi &= g, \text{ on } \Gm.
\end{split}
\end{equation}

\begin{remark}\label{rem:choice_legendre_transform}
If we had chosen a different form for \eqref{eqn:modified_energy_for_fixed_pot}, i.e. replace $\iG \sigma g$ with something else, then we would have derived a \emph{different PDE} than \eqref{eqn:electro-static_eqn_fixed_pot_final}.  Since we know \emph{a priori} the PDE that should be satisfied (with boundary potential fixed), the above argument justifies the form of \eqref{eqn:modified_energy_for_fixed_pot}.
\end{remark}

\subsubsection{A New Thermo-dynamic Potential}\label{sec:new_thermo_potential}

Going back to \eqref{eqn:modified_energy_for_fixed_pot}, \eqref{eqn:electro-static_eqn_energy_fixed_pot} and using \eqref{eqn:electro-static_eqn_fixed_charge_final}, we find that
\begin{equation}\label{eqn:electro-static_eqn_energy_fixed_pot_thermo_potential}
\begin{split}
\widehat{U}(g) &= \frac{1}{2} \iO \varepsilon_0 \varepsilon_r |\nabla \varphi|^2 - \iG \vn \cdot (\varepsilon_0 \varepsilon_r \nabla \varphi - \vP) \underbrace{g}_{=\varphi},
\end{split}
\end{equation}
where $\varphi \equiv \varphi (g)$ solves \eqref{eqn:electro-static_eqn_fixed_pot_final}.  Next, using \eqref{eqn:electro-static_eqn_fixed_pot_final} and integration by parts, we can see the minus sign coming in.  The final result is
\begin{equation}\label{eqn:electro-static_eqn_fixed_pot_dual_energy}
\begin{split}
\widehat{U}(g) &= \frac{1}{2} \iO \varepsilon_0 \varepsilon_r |\nabla \varphi|^2 + \iO - \nabla \cdot (\varepsilon_0 \varepsilon_r \nabla \varphi - \vP) \varphi - \iO (\varepsilon_0 \varepsilon_r \nabla \varphi - \vP) \cdot \nabla \varphi \\
&= -\frac{1}{2} \iO \varepsilon_0 \varepsilon_r |\nabla \varphi|^2  + \iO \vP \cdot \nabla \varphi + \iO \rho_{B} \varphi.
\end{split}
\end{equation}

\begin{remark}
Define the following functional $J(\varphi)$ for all $\varphi$ (sufficiently smooth) such that $\varphi |_{\Gm} = g$:
\begin{equation}\label{eqn:electro-static_PDE_functional}
\begin{split}
J(\varphi) &= \frac{1}{2} \iO \varepsilon_0 \varepsilon_r |\nabla \varphi|^2 - \iO \vP \cdot \nabla \varphi - \iO \rho_{B} \varphi,
\end{split}
\end{equation}
where $\vP$ and $\rho_{B}$ are treated as given functions, and note that $\widehat{U}(g) = -J(\varphi (g))$.

Computing the variational derivative, we get
\begin{equation}\label{eqn:electro-static_PDE_functional_deriv_pt1}
\begin{split}
\delta_{\varphi} J(\varphi;\eta) &= \iO \varepsilon_0 \varepsilon_r \nabla \varphi \cdot \nabla \eta - \iO \vP \cdot \nabla \eta - \iO \rho_{B} \eta,
\end{split}
\end{equation}
where $\eta |_{\Gm} = 0$ (i.e. we cannot change the value of $\varphi$ on $\Gm$).  Integrating by parts, and setting equal to zero, we get
\begin{equation}\label{eqn:electro-static_PDE_functional_deriv_pt2}
\begin{split}
\delta_{\varphi} J(\varphi;\eta) &= \iO \left[ -\nabla \cdot \left( \varepsilon_0 \varepsilon_r \nabla \varphi - \vP \right) - \rho_{B} \right] \eta = 0,
\end{split}
\end{equation}
for all $\eta$ that vanish on $\Gm$.  The only possibility for \eqref{eqn:electro-static_PDE_functional_deriv_pt2} to be true is that $\varphi$ solves \eqref{eqn:electro-static_eqn_fixed_pot_final}.

Therefore, it seems that the correct thermodynamic potential to use (when the boundary voltage is fixed) is always given by $-J(\varphi)$, where $J$ is the convex functional whose Euler-Lagrange equation is the electro-static PDE, and $\varphi$ \emph{solves the electro-static PDE}.
\end{remark}

We conclude by noting that $\widehat{U}$ depends on $g$, $\varepsilon_r$, $\vP$, and the domain $\Om$.  There is no dependence on $\varphi$ because $\varphi$ solves \eqref{eqn:electro-static_eqn_fixed_pot_final}, which depends on $g$, $\varepsilon_r$, $\vP$, and the domain $\Om$.

\section{Fixing a General Boundary Quantity}\label{sec:fixing_general_bdy_quantity}

In order to better motivate the advantage of the modeling approach above, we will fix a ``general'' function of charge.  The modified energy \eqref{eqn:modified_energy_for_fixed_pot} is similar to before:
\begin{equation}\label{eqn:modified_energy_for_fixed_quant}
	\cA(\sigma) = U(\sigma) - \iG \Psi (\sigma),
\end{equation}
where $\Psi(\sigma)$ is the energy (density) of the charge $\sigma$ on the boundary.  Note that $\sigma \equiv (\vn \cdot \varepsilon_0 \varepsilon_r \nabla \varphi) - \vP \cdot \vn$ (boundary charge).



So, we want to minimize:
\begin{equation}\label{eqn:electro-static_eqn_energy_fixed_quant}
\begin{split}
\widehat{U} &= \min_{\sigma} \cA(\sigma),
\end{split}
\end{equation}
where $\widehat{U}$ depends on $\Psi$, and $\varphi = \varphi(\sigma)$ solves \eqref{eqn:electro-static_eqn_fixed_charge_final}, and recall \eqref{eqn:electro-static_eqn_fixed_charge_perturb}.

Therefore, at equilibrium, we must have
\begin{equation}\label{eqn:electro-static_eqn_energy_fixed_quant_min_prop}
\begin{split}
	\delta_{\sigma} \cA(\sigma; \xi) &= 0, \quad \forall \xi.
\end{split}
\end{equation}
We assume this equilibrium point is indeed a minimizer because $U(\sigma)$ \emph{is convex} in $\sigma$ and we assume the term $\Psi(\sigma)$ does not destroy this.  Let $\hat{\sigma}$ be the unique solution of \eqref{eqn:electro-static_eqn_energy_fixed_quant_min_prop}.

Computing \eqref{eqn:electro-static_eqn_energy_fixed_quant_min_prop} explicitly, we get
\begin{equation}\label{eqn:electro-static_eqn_energy_fixed_quant_E-L_pt_1}
\begin{split}
\iO \varepsilon_0 \varepsilon_r \nabla \varphi (\hat{\sigma}) \cdot \nabla \varphi' - \iG \Psi'(\hat{\sigma}) \xi &= 0, \quad \forall \xi.
\end{split}
\end{equation}
Now do integration by parts as we did before,
\begin{equation}\label{eqn:electro-static_eqn_energy_fixed_quant_E-L_pt_2}
\begin{split}
\iO \underbrace{-\nabla \cdot \left( \varepsilon_0 \varepsilon_r \nabla \varphi' \right)}_{=0} \varphi (\hat{\sigma}) + \iG \underbrace{\vn \cdot \left( \varepsilon_0 \varepsilon_r \nabla \varphi' \right)}_{=\xi} \varphi (\hat{\sigma}) - \iG \Psi'(\hat{\sigma}) \xi = 0, \quad \forall \xi,
\end{split}
\end{equation}
and so
\begin{equation}\label{eqn:electro-static_eqn_energy_fixed_quant_E-L_pt_3}
\begin{split}
	\iG \left( \varphi (\hat{\sigma}) - \Psi'(\hat{\sigma}) \right) \xi = 0, \quad \forall \xi.
\end{split}
\end{equation}
This implies that $\varphi (\hat{\sigma})$ solves
\begin{equation}\label{eqn:electro-static_eqn_fixed_quant_final}
\begin{split}
-\nabla \cdot (\varepsilon_0 \varepsilon_r \nabla \varphi) &= \rho_B - \nabla \cdot \vP, \text{ in } \Om, \\
\varphi - \Psi' \left( \vn \cdot \varepsilon_0 \varepsilon_r \nabla \varphi - \vP \cdot \vn \right) &= 0, \text{ on } \Gm,
\end{split}
\end{equation}
whose boundary condition is a kind of non-linear Robin condition.

\textbf{Example.}  Suppose $\Psi(\sigma) = \sigma g$, where $g$ is a given fixed function. Then, the boundary condition reduces to
\begin{equation*}
	\varphi = g, \text{ on } \Gm,
\end{equation*}
which was the case we presented in Section \ref{sec:general_arg}.

\textbf{Example.}  Suppose $\Psi(\sigma) = -\sigma^2 / 2$. In this case, the boundary condition reduces to
\begin{equation*}
\varphi + \vn \cdot \varepsilon_0 \varepsilon_r \nabla \varphi = \vP \cdot \vn, \text{ on } \Gm,
\end{equation*}
which is a standard linear Robin boundary condition.  Note that the choice of $\Psi$ here ensures that \eqref{eqn:modified_energy_for_fixed_quant} is convex in $\sigma$.

\section{Conclusion}\label{sec:conclusion}

We have presented a complete derivation of electrical energy for a distributed dielectric system.  It is our hope that this discussion has demystified the ``minus sign'' in the electrical energy.

\newpage


\bibliographystyle{abbrv}
\bibliography{MasterBibTeX}

\begin{thebibliography}{1}

\bibitem{Feynman_Lectures1964}
R.~P. Feynman, R.~B. Leighton, and M.~Sands.
\newblock {\em The Feynman Lectures on Physics}.
\newblock Addison-Wesley Publishing Company, 1964.

\bibitem{Landau_Book1960}
L.~D. Landau and E.~M. Lifshitz.
\newblock {\em Electrodynamics of Continuous Media}, volume~8 of {\em Course of
  Theoretical Physics}.
\newblock Addison-Wesley, 1960.

\bibitem{Onuki_inbook2005}
A.~Onuki.
\newblock Electric field effects near critical points.
\newblock In S.~Rzoska and V.~Zhelezny, editors, {\em Nonlinear Dielectric
  Phenomena in Complex Liquids}, volume 157 of {\em NATO Science Series II:
  Mathematics, Physics and Chemistry}, pages 113--141. Springer Netherlands,
  2005.

\end{thebibliography}

\end{document}